\documentclass[12pt]{article}
\pdfoutput=1
\usepackage{a4wide,epsfig,psfrag,amsmath,amssymb,cite,scalefnt}
\usepackage{color}

\graphicspath{ {figs/} }

\newcommand{\ep}{\epsilon}

\allowdisplaybreaks
\newcommand{\Li}{\text{Li}}
\newcommand{\N}{\nonumber}

\newcommand{\gsim}{\;\rlap{\lower 3.5 pt \hbox{$\mathchar \sim$}} \raise 1pt
 \hbox {$>$}\;}
\newcommand{\lsim}{\;\rlap{\lower 3.5 pt \hbox{$\mathchar \sim$}} \raise 1pt
 \hbox {$<$}\;}

\begin{document}

\title{\vskip-3cm{\baselineskip14pt
    \begin{flushleft}
      \normalsize TTP15-019, DESY 15-090, NIKHEF-2015-022
% LPN14-101
  \end{flushleft}}
  \vskip1.5cm
  Exact N$^3$LO results for $q q^\prime\to H +X$
}

\author{
  Chihaya Anzai$^{(a)}$,
  Alexander Hasselhuhn$^{(a)}$,
  Maik H\"oschele$^{(a)}$,\\
  Jens Hoff$^{(b)}$,
  William Kilgore$^{(c)}$,
  Matthias Steinhauser$^{(a)}$,
  Takahiro Ueda$^{(d)}$
  \\[1em]
  {\small\it (a) Institut f{\"u}r Theoretische Teilchenphysik, 
    Karlsruhe Institute of Technology (KIT)}\\
  {\small\it 76128 Karlsruhe, Germany}
  \\[1em]
  {\small\it (b) Deutsches Elektronen Synchrotron DESY, Platanenallee 6, 15738
  Zeuthen, Germany}
  \\[1em]
  {\small\it (c) Physics Department, Brookhaven National Laboratory, Upton, New
    York 11973, USA}
  \\[1em]
  {\small\it (d) Nikhef Theory Group, Science Park 105, 1098 XG Amsterdam, The
  Netherlands}
}

\date{}

\maketitle

\thispagestyle{empty}

\begin{abstract}

  We compute the contribution to the total cross section for the inclusive
  production of a Standard Model Higgs boson induced by two quarks with
  different flavour in the initial state. Our calculation is exact in the
  Higgs boson mass and the partonic center-of-mass energy. We describe the
  reduction to master integrals, the construction of a canonical basis, and
  the solution of the corresponding differential equations.  Our analytic
  result contains both Harmonic Polylogarithms and iterated integrals with
  additional letters in the alphabet.

\medskip


\end{abstract}

\thispagestyle{empty}

%- }}}

\newpage

%- {{{ Introduction:

\section{Introduction}

The precise determination of the properties of the recently discovered Higgs
boson~\cite{:2012gk,:2012gu} is among the main tasks of the upcoming
run~II of the CERN Large Hadron Collider (LHC).  A crucial input to this
enterprise is the total production cross section in gluon fusion.

Leading order (LO) contributions to $\sigma(pp\to H+X)$ were already computed
by the end of the 1970s in
Refs.~\cite{Wilczek:1977zn,Ellis:1979jy,Georgi:1977gs,Rizzo:1979mf} and the
next-to-leading order (NLO) QCD corrections have been available for almost 20
years~\cite{Dawson:1990zj,Spira:1995rr} including the exact
dependence on the top quark mass (see also Ref.~\cite{Harlander:2005rq} for
analytic results of the virtual corrections). NLO electroweak corrections have
been computed in Ref.~\cite{Actis:2008ug} and mixed QCD-electroweak
corrections are considered in Ref.~\cite{Anastasiou:2008tj}.

At LHC energies the NLO QCD corrections amount to 80-100\% of the LO
contributions which makes it mandatory to compute higher-order perturbative
corrections.  At the beginning of the century three groups independently
evaluated the next-to-next-to-leading order (NNLO)
corrections~\cite{Harlander:2000mg,Harlander:2002wh,Anastasiou:2002yz,Ravindran:2003um}
in the limit of an infinitely heavy top quark. Finite top quark mass effects,
which have been investigated in
Refs.~\cite{Marzani:2008az,Harlander:2009bw,Pak:2009bx,Harlander:2009mq,Pak:2009dg,Harlander:2009my,Pak:2011hs},
turn out to be at most of the order of 1\%.

At next-to-next-to-next-to-leading order
(N$^3$LO) several groups have contributed building blocks to the total cross
section.  In Refs.~\cite{Chetyrkin:1997un,Schroder:2005hy,Chetyrkin:2005ia}
the effective Higgs-gluon coupling has been computed to four-loop
accuracy. In preparation of the N$^3$LO calculations the
${\cal O}(\epsilon)$ contributions to the NNLO master integrals have been
computed in Refs.~\cite{Pak:2011hs,Anastasiou:2012kq} where $d=4-2\epsilon$ is
the number of
space-time dimensions in dimensional regularization. Results for the LO,
NLO and NNLO partonic cross sections expanded up to order $\epsilon^3$,
$\epsilon^2$ and $\epsilon^1$, respectively, have been published in
Refs.~\cite{Hoschele:2012xc,Buehler:2013fha}.  All contributions from
convolutions of partonic cross sections with splitting functions, which are
needed for the complete N$^3$LO calculation, are provided in
Refs.~\cite{Hoschele:2012xc,Hoeschele:2013gga,Buehler:2013fha}.  The full
scale-dependence of the N$^3$LO expression has been constructed in
Ref.~\cite{Buehler:2013fha}.  Three-loop ultraviolet counterterms needed
for $\alpha_s$~\cite{Tarasov:1980au,Larin:1993tp} and the
operator in the effective Lagrangian~\cite{Spiridonov:1984br}
were computed long ago.

Within the effective theory, three-loop virtual corrections to the Higgs-gluon
form factor have been obtained by two independent
calculations~\cite{Baikov:2009bg,Gehrmann:2010ue} (see also
Ref.~\cite{Lee:2010cga}).  The single-soft current to two-loop order has been
computed in Refs.~\cite{Duhr:2013msa,Li:2013lsa} which is an important
ingredient to the two-loop corrections with one additional real radiation. The
latter have been computed in Refs.~\cite{Dulat:2014mda,Duhr:2014nda}.  The
single-real radiation contribution which originates from the square of
one-loop amplitudes has been computed exactly in terms of the Higgs boson
  mass and the partonic center-of-mass energy in
Refs.~\cite{Anastasiou:2013mca,Kilgore:2013gba}.  The soft limit of the phase
space integrals for Higgs boson production in association with two soft
partons were computed in Refs.~\cite{Anastasiou:2014vaa,Li:2014bfa}, in the
latter reference even to all orders in $\epsilon$.  The triple-real
contribution to the gluon-induced partonic cross section has been considered
in Ref.~\cite{Anastasiou:2013srw} in the soft limit.  In particular, a method
has been developed which allows the expansion around the soft limit.  A
similar analysis for the double-real-virtual contributions has been published
in Ref.~\cite{Anastasiou:2015yha}.

The two leading terms in the threshold expansion for the complete N$^3$LO total
Higgs production cross section through gluon fusion have been presented in
Refs.~\cite{Anastasiou:2014vaa,Anastasiou:2014lda,Li:2014afw}.  However, for
physical applications more terms in the threshold expansion are
necessary~\cite{Anastasiou:2014lda}. In fact, in
Ref.~\cite{Anastasiou:2015ema} more than
30 expansion terms have been computed which is
sufficient for all phenomenological applications.  It is important to
cross-check the result of Ref.~\cite{Anastasiou:2015ema}.  In this paper we
present the first step in this direction. In particular, results are obtained
which are exact in the Higgs boson mass and the partonic center-of-mass energy.

Further activities concern the development of systematic approaches to 
compute the master integrals for $\sigma(pp\to H+X)$, see, e.g.,
Refs.~\cite{Anastasiou:2013srw,Anastasiou:2013mca,Kilgore:2013gba,Hoschele:2014qsa,Dulat:2014mda}. 

Several groups have constructed approximate N$^3$LO results for the total
cross section taking into account information from the soft-gluon
approximation and the high-energy
limit~\cite{Moch:2005ky,Ahrens:2010rs,Ball:2013bra,Bonvini:2014jma,Bonvini:2014joa,Catani:2014uta,deFlorian:2014vta}. 

In the following, we briefly outline the framework
which we use for our calculation. In the limit of
an infinitely heavy top quark the effective interaction
of the Higgs boson with gluons is described by the Lagrange density
\begin{eqnarray}
  {\cal L}_{Y,\rm eff} &=& -\frac{H^0}{4v^0} C_1^0  (G_{\mu\nu} G^{\mu\nu})^0 
  + {\cal L}_{QCD}^{(5)}
  \,,
  \label{eq::leff}
\end{eqnarray}
where ${\cal L}_{QCD}^{(5)}$ is the usual QCD Lagrange density with five
massless quarks, $H$ denotes the Higgs field, $v$ its vacuum expectation
value and $C_1$ is the matching coefficient between the full and the
effective theory.  $G_{\mu\nu}$ is the gluonic field strength tensor
constructed from fields and couplings already present in ${\cal
  L}_{QCD}^{(5)}$. The superscript ``0'' denotes the bare quantities. Note
that the counterterms of $H^0/v^0$ are of higher order in the electroweak
coupling constants.

The top quark mass enters the cross section via the matching coefficient
$C_1$ whereas the quantities in the effective theory depend on
\begin{eqnarray}
  x &=& \frac{m_h^2}{\hat{s}} \,,
\end{eqnarray}
where $m_h$ is the Higgs boson mass and
$\sqrt{\hat{s}}$ the partonic center-of-mass energy.
For later convenience we also introduce the variable
\begin{eqnarray}
  y &=& 1-x\,.
\end{eqnarray}

At the partonic level several sub-processes initiated by quarks and gluons in
the initial state have to be considered. The 
numerically most important but also technically most complicated 
contribution is the one with two gluons in the initial state.
In the present paper we consider the subprocess $q q^\prime\to H+X$ at NNLO
and N$^3$LO.
Its phenomenological impact is very small, but we use this
process to demonstrate our method which leads to exact results
in $x$ and avoids the high-order soft expansion.

\begin{figure}[t]
  \begin{center}
    \includegraphics[width=0.18\textwidth]{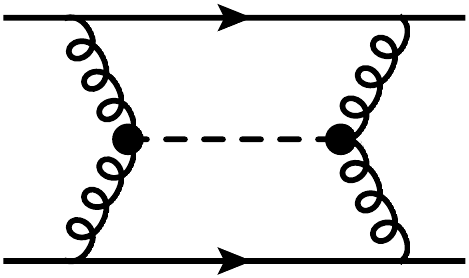} \hfill
    \includegraphics[width=0.18\textwidth]{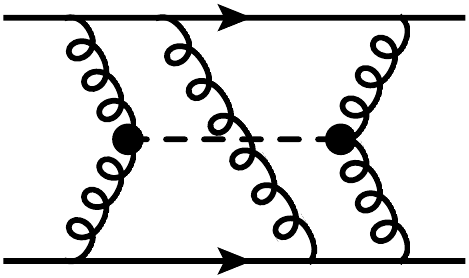} \hfill
    \includegraphics[width=0.18\textwidth]{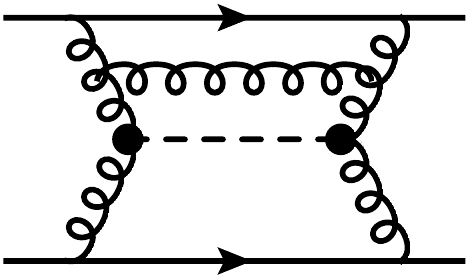} \hfill
    \includegraphics[width=0.18\textwidth]{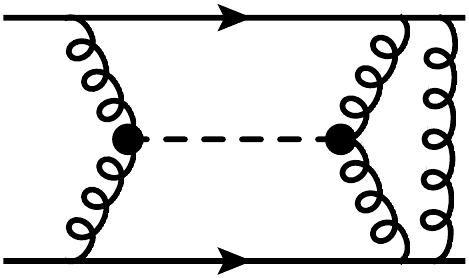} \hfill
    \includegraphics[width=0.18\textwidth]{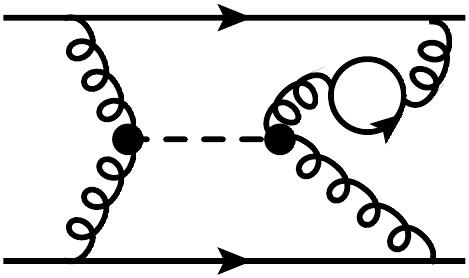}
    \\
    \caption{\label{fig::qp2qp} Sample Feynman diagrams for $q q^\prime \to q
      q^\prime$.  The imaginary parts due to Higgs boson cuts provide the
      cross section for the process $q q^\prime \to H + X$ at NNLO and
      N$^3$LO.  Solid, curly and dashed lines represent quarks, gluons and
      Higgs bosons, respectively and blobs denote the effective
      Higgs-gluon couplings.}
  \end{center}
\end{figure}

For the calculation of the total cross section it is convenient to consider
the imaginary part of the forward scattering amplitude $q q^\prime \to q
q^\prime$.  Sample Feynman diagrams contributing at NNLO and N$^3$LO are shown
in Fig.~\ref{fig::qp2qp}. To obtain the cross section all cuts involving the
Higgs boson have to be computed which means that both three- and four-particle
cuts have to be considered at N$^3$LO. There are no two-particle cuts.

The remainder of the paper is organized as follows: In the next Section we
discuss the reduction of the full set of integrals to master integrals and the
construction of the canonical basis.  For the latter integrals a system of
differential equations is derived. The following two sections are dedicated to
the evaluation of the initial conditions involving cuts of three
(Section~\ref{sec::3p}) and four (Section~\ref{sec::4p}) particles.  In
Section~\ref{sec::iter} we introduce recursively defined iterated integrals
which are needed for the analytic representation of the final result. The
partonic cross section is discussed in Section~\ref{sec::res} where
analytic results are given. Finally we conclude in Section~\ref{sec::con}.

%- }}}
%- {{{ Reduction, canon. integrals:

\section{Reduction and canonical master integrals}

\begin{figure}[t]
  \begin{center}
    \includegraphics[width=.8\textwidth]{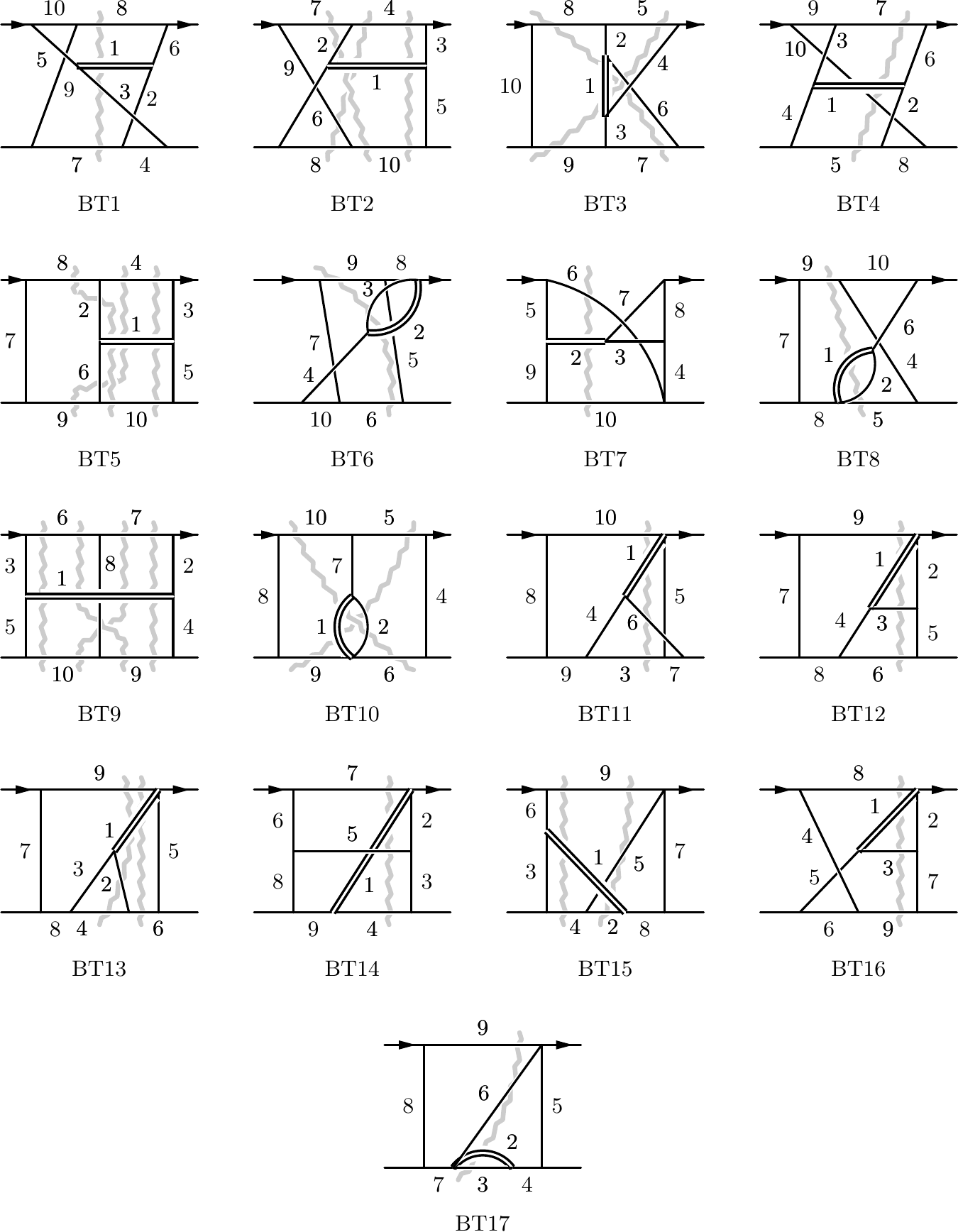}
    \\
    \caption{\label{fig::int_fam} Graphical representations of the 17
      three-loop integral families. Plain and double lines indicate massless 
      propagators and the Higgs boson lines, respectively, 
      and the wavy lines indicate the
      possible cuts. }
  \end{center}
\end{figure}

We generate all two- and three-loop forward-scattering amplitudes for the
process $q(p_1) q^\prime(p_2) \to q(p_1) q^\prime(p_2)$ involving a virtual
Higgs boson with the help of {\tt qgraf}~\cite{Nogueira:1991ex} and process
the output file to select the contributions which contain cuts through the
Higgs boson line.  This leads to 1 two-loop and 224 three-loop Feynman
diagrams. At three-loop order the corresponding amplitudes can be mapped to 17
integral families which are shown in Fig.~\ref{fig::int_fam}. For each of them
reduction tables are generated using a combination of the publicly available
program {\tt FIRE}~\cite{Smirnov:2014hma} and in-house programs, {\tt rows}
and {\tt TopoID}~\cite{Grigo:2014oqa}, which implement the Laporta
algorithm~\cite{Laporta:2001dd}. The use of {\tt rows} and {\tt TopoID}
guarantees that all available symmetries are exploited which is important to
minimize the number of master integrals. After completing the reduction for
each family we obtain 332 master integrals. In our next step we minimize
the number of integrals by simultaneously considering all families which
leaves us with 111 master integrals, 108 of which are needed for the cross
section.  In the following we refer to this set of master integrals as 
  ``Laporta master integrals''.

Note that we have performed the calculation for general gauge
parameter $\xi$ which drops out after relating master integrals
from the different families. This constitutes a strong check on the
correctness of our result.

For the evaluation of the master integrals we follow the ideas of
Ref.~\cite{Henn:2013pwa} and construct a canonical basis which allows for a
simple and straightforward solution of the corresponding differential
equations (see Refs.~\cite{Argeri:2007up,Smirnov:2012gma} for reviews on the
use of differential equations for the computation of Feynman integrals).
Whereas most of our calculation is automated to a high degree the construction
of the canonical basis requires manual manipulations of each individual
integral.  We have applied several tricks described in the
literature~\cite{Cachazo:2008vp,ArkaniHamed:2010gh,Henn:2013wfa,Henn:2013tua,Henn:2014qga} and also follow the algorithm developed in
Ref.~\cite{Hoschele:2014qsa} which allows the construction of canonical 
master integrals in coupled subsystems.
In Ref.~\cite{Lee:2014ioa} an algorithm has been suggested which automates the
construction of the canonical basis. However, a public implementation
is not yet available.

In a canonical basis the differential equations have the form
\begin{eqnarray}
  \partial_x f(x,\epsilon) &=& \epsilon A(x) f(x,\epsilon)
  \,,
\end{eqnarray}
where $f(x,\epsilon)$ is a vector containing all canonical master integrals.
In our case the matrix $A(x)$ can be written as
\begin{eqnarray}
  A(x) &=& \frac{a}{1-x} + \frac{b}{1+x} + \frac{c}{x} + \frac{d}{1+4x} +
  \frac{e}{x\sqrt{1+4x}}
  \,,
  \label{eq::Ax}
\end{eqnarray}
where $a,\ldots,e$ are constant matrices.  The first three terms on
the right-hand side of Eq.~(\ref{eq::Ax}) lead to the well-known
Harmonic Polylogarithms (HPLs)~\cite{Remiddi:1999ew} (see
Refs.~\cite{Maitre:2005uu,Maitre:2007kp} for a convenient {\tt
  Mathematica} implementation) in the solution of the master
integrals.  The fourth and fifth terms in Eq.~(\ref{eq::Ax}) are only
needed for the integral family \text{BT3} as we will describe in
detail in Section~\ref{sec::iter}.

Besides the simple solution of the differential equations the canonical basis
also has the advantage that for the initial conditions only the leading terms
of order $y^0$ are needed in the soft limit. As a consequence, no explicit
calculation is needed in case the first non-zero contribution of a canonical
master integral is of ${\cal O}(y)$ or higher.  In our calculation the
boundary conditions are computed for Laporta master integrals.  Afterwards the
results are transformed to the canonical basis.

%- }}}
%- {{{ 3-particle cuts:

\section{\label{sec::3p}Three-particle cuts}

The three-particle-cut contributions contain a one-loop sub-diagram.  As our
first step we represent the loop in terms of Mellin-Barnes integrals and
perform the momentum integration. Afterwards in the soft limit
all integrals are represented as
phase space integrals of soft partons, which can be converted to integrals
over energies and angles.  These integrals are also calculable using
Mellin-Barnes integrals.  Hence, we obtain multifold Mellin-Barnes integral
representations for each master integral in the soft limit.
They are evaluated extending the
method developed in Ref.~\cite{Anzai:2012xw} for the calculation of the
three-loop static potential.  The notation is mainly adopted from
Ref.~\cite{Anastasiou:2013srw} where four-particle cuts have been considered.
In this reference also a technique has been developed which transforms soft
phase-space integrals to Mellin-Barnes integrals, which has been applied in
Ref.~\cite{Anastasiou:2015yha} to three-particle phase space contributions.
In contrast to Ref.~\cite{Anastasiou:2015yha} we do not apply the method of
regions to compute the integrals.

Before describing the procedure in more detail we have to introduce some
notation. We denote the external momenta by $p_{1}, p_{2}$ and the light
momenta involved in the cut by $p_{3}, p_4$ and $p_5$, where $p_5$ will occur
for the four-particle phase space integrals in Section~\ref{sec::4p}.  Loop
momenta are denoted by $v_{i}$.  According to Ref.~\cite{Anastasiou:2013srw}
the scaling of the phase space momenta in the soft limit is given by $p_i\sim
\sqrt{s}$ for $i=1,2$ and $p_i\sim y\sqrt{s}$ for $i=3,4$ and $5$ in the
center-of-mass frame of the incoming quarks. We eliminate the momentum of the
Higgs boson in favour of the momenta of the massless partons and
define rescaled scalar products
\begin{eqnarray}
  s_{ij} &=&
  \begin{cases} 
    \frac{(p_i-p_j)^2}{sy}, & i=1,2 \text{ and } j>2\,,\\
    \frac{(p_i+p_j)^2}{sy^2} & i>2 \text{ and } j>2\,.
  \end{cases}
  \label{eq:sij}
\end{eqnarray}
Furthermore, we use the energies and angles parametrization
\begin{eqnarray}
  \frac{p_1}{\sqrt{s}} &=&
  \frac{1}{2}\beta_1=\frac{1}{2}(1,0_{d-2},1)^\mathsf{T}\,,
  \nonumber\\ 
  \frac{p_2}{\sqrt{s}} &=&
  \frac{1}{2}\beta_2=\frac{1}{2}(1,0_{d-2},-1)^\mathsf{T}\,,
  \nonumber\\ 
  \frac{p_i}{\sqrt{s}y} &=& \frac{1}{2}E_i\beta_i \quad\mbox{for}\quad i>2\,,
  \label{eq:pi}
\end{eqnarray}
where $E_i$ parametrize the partons' energies and $\beta_i$ their
$d$-dimensional velocities.  $0_{d-2}$ is an abbreviation for a sequence of
$d-2$ zeros. For later convenience we also introduce
$\beta_{ij}=\beta_i\cdot\beta_j$. 

In the following we exemplify the
individual steps of the algorithm on the integral
\begin{eqnarray}
  B_9 &=& \mathrm{BT9}(1,1,1,0,1,0,1,0,1,1,0,0)
  \nonumber \\
  &=& \int {\rm d}\Phi_{3}^{s}\int\frac{ {\rm d}^{
      d}v}{\left(2\pi\right)^{{d}}}
  \frac{N}
  {v^{2}(p_{1}{-}p_{3})^{2}(p_{1} {-} p_{3} {-}
    p_{4}+v)^{2}(p_{1}+p_{2}{-}p_{3}{-}p_{4}+v)^{2}}
  \,.
\end{eqnarray}
$N$ is a normalization factor given by
\begin{eqnarray}
  N &=& \frac{1}{2\pi} 
  \left( \frac{ (4\pi)^{2-\epsilon} }{\Gamma(1+\epsilon)} \right)^3
  \,,
\end{eqnarray}
where the factors $\Gamma(1+\epsilon)$ and $(4\pi)^\epsilon$ are introduced for
convenience
and ${\rm d}\Phi_{3}^{s}$ is the soft three-particle phase space measure
which can be written as
\begin{eqnarray}
  \int {\rm d}\Phi_{3}^{s} &=& 
  (2\pi)^{-5+4\epsilon}
  2^{-6+4 \epsilon}
  s^{1-2\epsilon}y^{3-4\epsilon}
  \delta
  \left(1-\sum_{i=3}^{4}E_{i}\right)\prod_{i=3}^{4}E_{i}^{1-2\epsilon}
  \int {\rm d}E_{i}\int {\rm d}\Omega_{i}^{d-1} 
  \,.
  \label{eq::phi3}
\end{eqnarray}
$\Omega_{i}^{d-1}$ is the $d$-dimensional solid angle.

The algorithm for the computation of the three-particle cut contribution is 
as follows:
\begin{enumerate}
\item Introduce a regularization parameter $\delta$ for the numerators.
  This is necessary to avoid terms $\Gamma(0)$
  which otherwise could appear in step~\ref{enu:E_Omega} below.
  We introduce $\delta$ to the exponent of the scalar products, namely,
  $\left(p_{i}+\cdots\right)^{2} \rightarrow
  \lim_{\delta\rightarrow0}\left[\left(p_{i}+\cdots\right)^{2}\right]^{1+\delta}$.

\item Perform subloop integration and introduce Mellin-Barnes integrals.
  \label{enu:Perform-subloop-integration}

  We (i) introduce Feynman parameters to combine propagators
  involving loop momenta, (ii) perform loop integration and (iii) introduce
  Mellin-Barnes variables to obtain a factorization of the Feynman 
  variables~\cite{Anastasiou:2013srw} using the formula
  \begin{eqnarray}
    \label{eq:repDenomMB}
    \frac{1}{(X+Y)^\lambda} &=& 
    \frac{1}{\Gamma(\lambda)}\frac{1}{2\pi i}
    \int_{-i\infty}^{+i\infty} {\rm d}z\, 
    \Gamma(\lambda+z)\Gamma(-z)\frac{Y^z}{X^{\lambda+z}}
    \,.
  \end{eqnarray}
  In our example we obtain a one-fold
  Mellin-Barnes integral over $z_1$ which has the following form
  \begin{eqnarray}
    B_9 &=&
    \int {\rm d}\Phi_{3}^{s}\int\frac{{\rm d}z_{1}}{2\pi i} \times
    \nonumber\\&&
    \frac{2^{7-4\epsilon}\pi^{3-2\epsilon}
      \Gamma(-z_{1})\Gamma(z_{1}+1)\Gamma(-\epsilon)\Gamma(-z_{1}-\epsilon) 
      \Gamma(z_{1}+\epsilon+1)}{(p_{1}{-}p_{3})^{2}
      (p_{1}+p_{2} {-}p_{3} {-}p_{4})^{-2z_{1}}(p_{1}{-}p_{3}{-}p_{4})^{2z_{1}+2\epsilon+2}
      \Gamma(1-2\epsilon)\Gamma^3(\epsilon+1)}  
    \,.
    \nonumber\\
  \end{eqnarray}
  
\item Express the propagators in terms of velocities and energies, and take the 
  soft limit, i.e., $y\to0$.

  Using Eq.~(\ref{eq:sij}) we can replace the propagators
  in our examples $B_9$ as
  \begin{eqnarray}
    \frac{1}{(p_{1}{-}p_{3})^{2}} & \rightarrow &
    {-}\frac{2}{syE_{3}\beta_{13}}\,, \nonumber\\ 
    \frac{1}{(p_{1}{-}p_{3}{-}p_{4})^{2+2z_{1}+2\epsilon}}
    & \rightarrow &
    \left(-\frac{1}{2}syE_{3}\beta_{13}-\frac{1}{2}syE_{4}\beta_{14}+\frac{1}{2}sy^{2}E_{3}E_{4}\beta_{34}\right)^{-z_{1}-\epsilon-1}\,,
    \nonumber\\ 
    \frac{1}{(p_{1}+p_{2}{-}p_{3}{-}p_{4})^{-2z_{1}}}
    & \rightarrow &
    \left(\frac{1}{2}s {\beta_{12}}-\frac{1}{2}syE_{3}\beta_{13}
    -\frac{1}{2}syE_{4}\beta_{14}-\frac{1}{2}syE_{3}\beta_{23} 
    \right.
    \nonumber\\&&\mbox{}
    \left.
    -\frac{1}{2}syE_{4}\beta_{24}
    +\frac{1}{2}sy^{2}E_{3}E_{4}\beta_{34}\right)^{z_{1}}\,.
  \end{eqnarray}
  To leading order in $y$ this becomes
  \begin{eqnarray}
    \frac{1}{(p_{1}{-}p_{3})^{2}} & \rightarrow &
    {-}\frac{2}{syE_{3}\beta_{13}}\,, \nonumber\\ 
    \frac{1}{(p_{1}{-}p_{3}{-}p_{4})^{2+2z_{1}+2\epsilon}}
    & \rightarrow &
    \left(-\frac{1}{2}syE_{3}\beta_{13}
    -\frac{1}{2}syE_{4}\beta_{14}\right)^{-z_{1}-\epsilon-1} 
    \,, \nonumber\\ 
    \frac{1}{(p_{1}+p_{2}{-}p_{3}{-}p_{4})^{-2z_{1}}}
    & \rightarrow & \left(\frac{1}{2}s {\beta_{12}}\right)^{z_{1}} \,.
    \label{eq::pi-betaE}
  \end{eqnarray}

\item Introduce Mellin-Barnes variables to factor the $\beta_{ij}$ and
  $E$ dependence.

  In our example a further Mellin-Barnes parameter $z_{2}$ has to be introduced
  to decompose the sum on the r.h.s of Eq.~(\ref{eq::pi-betaE}).
  Afterwards, the energy integrations are trivial and we obtain
  \begin{eqnarray}
    B_9 &=&
    \frac{2^{5\epsilon-2}\pi^{2\epsilon-2}s^{-3\epsilon-1}
      \Gamma(-\epsilon)}{\Gamma(1-2\epsilon)
      \Gamma^3(\epsilon+1)}\int\frac{{\rm d}z_{1}}
         {2\pi i}\int\frac{{\rm d}z_{2}}{2\pi i}(-1)^{z_{1}}y^{-z_{1}-5\epsilon+1}
         \Gamma(-z_{1})\Gamma(z_{1}+1)\Gamma(-z_{2})
         \nonumber\\&&\mbox{} \times
         \frac{\Gamma(-z_{1}-\epsilon)\Gamma(z_{2}-2\epsilon+2)
           \Gamma(-z_{1}-z_{2}-3\epsilon)\Gamma(z_{1}+z_{2}+\epsilon+1)}
              {\Gamma(-z_{1}-5\epsilon+2)}
              \nonumber\\&&\mbox{} \times
              \int {\rm d}\Omega_{3}^{{d-1}}
              \int {\rm d}\Omega_{4}^{{d-1}}
              \beta_{14}^{z_{2}}\beta_{13}^{-z_{1}-z_{2}-\epsilon-2}
              \,.
  \end{eqnarray}

\item \label{enu:E_Omega}
  Convert angular integrations to Mellin-Barnes integrations.
  This is achieved by using repeatedly~\cite{Somogyi:2011ir}
  \begin{eqnarray}
    &&
    \int\frac{{\rm d} 
      \Omega_{i}^{{d-1}}}
             {\beta_{j_{1}i}^{\alpha_{1}} \cdots\beta_{j_{n}i}^{\alpha_{n}}}
             \nonumber\\&&=
             \frac{2^{2-\sum_{m=1}^{n}
                 \alpha_{m}-2\epsilon}\pi^{1-\epsilon}}
                  {\prod_{k=1}^{n}{\Gamma}\left(\alpha_{k}\right)
                    \Gamma\left(2-\sum_{m=1}^{n}\alpha_{m}-2\epsilon\right)}
                  \Gamma\left(1-\sum_{m=1}^{n}\alpha_{m}-\epsilon
                  -\sum_{k=1}^{n}\sum_{l=k}^{n}z_{kl}\right)
                  \nonumber\\  
                  &&
                  \quad\times\int_{-i\infty}^{i\infty}
                  \left[\prod_{k=1}^{n}\prod_{l=k}^{n}\frac{{\rm d}z_{kl}}{2\pi i}
                    \Gamma\left(-z_{kl}\right)\beta_{j_{k}j_{l}}^{z_{kl}}\right]
                  \left[\prod_{k=1}^{n}\Gamma
                    \left(\alpha_{k}+\sum_{l=1}^{k}z_{lk}+\sum_{l=k}^{n}z_{kl}
                    \right)\right]
                  \,,
                  \label{eq::omega_int}
  \end{eqnarray}
  in order to perform the $\Omega$ integrations. 

  In the case of our example this leads to
  \begin{eqnarray}
    B_9 &=& 
    s^{-3\epsilon-1}
    \int\frac{{\rm d}z_{1}}{2\pi i}
    \int\frac{{\rm d}z_{2}}{2\pi i}y^{-z_{1}-5\epsilon+1}
    \cos(\pi z_{1})\Gamma(-z_{1})\Gamma(z_{1}+1)\Gamma(-z_{2})
    \nonumber\\&&\mbox{}
    \times\frac{\Gamma(-\epsilon)\Gamma(-z_{1}-\epsilon)
      \Gamma(z_{2}-\epsilon+1)\Gamma(-z_{1}-z_{2}-2\epsilon-1)
      \Gamma(z_{1}+z_{2}+\epsilon+1)}{\Gamma(1-2\epsilon)
      \Gamma^3(\epsilon+1)\Gamma(-z_{1}-5\epsilon+2)}
    \,.
    \nonumber\\
  \end{eqnarray}

\item Simplification with Barnes' Lemma.

  We use the routine \texttt{DoAllBarnes[]} of the package 
  \texttt{barnesroutines.m}~\cite{barnesroutines}.
  Before applying it, we convert the cosine to Gamma functions
  using either
  \begin{eqnarray}
    \cos(a) &=& 
    \frac{\psi^{(0)}\left(1-\frac{a}{\pi}\right)}
         {\Gamma\left(1-\frac{a}{\pi}\right)
           \Gamma\left(\frac{a}{\pi}\right)}-\frac{\psi^{(0)}
           \left(\frac{a}{\pi}\right)}
         {\Gamma\left(1-\frac{a}{\pi}\right)\Gamma\left(\frac{a}{\pi}\right)}
  \end{eqnarray}
  or
  \begin{eqnarray}
    \cos(a) &=&  
    \frac{\pi}{\Gamma\left(\frac{1}{2}-\frac{a}{\pi}\right)
      \Gamma\left(\frac{1}{2}+\frac{a}{\pi}\right)} 
  \end{eqnarray}
  depending on whether half-integer arguments are present in the 
  final expression or not. The latter should be avoided to arrive
  at simpler expressions.

\item Take the limit $\delta\rightarrow0$ (if needed) and expand in $y$
  and $\epsilon$.

  Using {\tt MBcontinue[]} from the package {\tt MB.m}~\cite{Czakon:2005rk},
  we can obtain Mellin-Barnes representations for the limits $\delta\to0$ and
  $\epsilon\to0$. To achieve this goal, we have slightly modified the code to
  prevent that $\log(y)$ terms appear.

  After that we expand the representation in $\delta$ and $\epsilon$ using
  {\tt MBexpand[]}, and in $y$ using {\tt MBasymptotics[]}~\cite{MBasymptotics}.

\item Further simplification of Mellin-Barnes integrals.

  We apply the following procedures iteratively:
  \begin{itemize}
  \item {\tt MBapplyBarnes[]}
  \item {\tt DoAllBarnes[]}
  \item Simplification of the integration contours such that all
    integrals with the same number of Mellin-Barnes parameters 
    have the same integration contours.
  \end{itemize}

\item Conversion to nested sums and their evaluation.

  To achieve this, we first use the residue theorem to convert the
  integrals to sums. In case this step generates divergent infinite
  sums, we introduce a regulator $e^{\pm\sigma c_{i}z_{i}}$ in the
  integrand, where the $c_{i}$'s are properly chosen numbers, $\sigma$
  is a regularization parameter, and the $z_{i}$'s are Mellin-Barnes
  parameters in the expression. For the evaluation of the sums, we use
  the summation program described in Ref.~\cite{Anzai:2012xw}.

\end{enumerate}

The final result for the integral $B_9$ reads
\begin{align}
\frac{B_{9}}{s^{-3\epsilon-1}}= & 
\frac{-\frac{1}{2}y^{2-5\epsilon}+\frac{1}{2}y^{2-4\epsilon}}{\epsilon^{3}}+\frac{-\frac{15}{4}y^{2-5\epsilon}+3y^{2-4\epsilon}}{\epsilon^{2}}+\frac{y^{2-5\epsilon}\left(\frac{11\zeta_{2}}{2}-\frac{175}{8}\right)+y^{2-4\epsilon}\left(14-6\zeta_{2}\right)}{\epsilon}\nonumber\\
  & 
+y^{2-5\epsilon}\left(\frac{165\zeta_{2}}{4}+18\zeta_{3}-\frac{1875}{16}\right)
+y^{2-4\epsilon}\left(-36\zeta_{2}-11\zeta_{3}+60\right)
\nonumber\\ & 
+\epsilon\left[y^{2-5\epsilon}\left(\frac{1925\zeta_{2}}{8}+135\zeta_{3}-\frac{31\zeta_{4}}{8}-\frac{19375}{32}\right)
\right.\nonumber\\&\left.\qquad
+y^{2-4\epsilon}\left(-168\zeta_{2}-66\zeta_{3}+\frac{105\zeta_{4}}{2}+248\right)\right]\nonumber\\
  & 
+\epsilon^{2}\bigg[y^{2-5\epsilon}\left(-198\zeta_{3}\zeta_{2}+\frac{20625\zeta_{2}}{16}+\frac{1575\zeta_{3}}{2}-\frac{465\zeta_{4}}{16}+294\zeta_{5}-\frac{196875}{64}\right)\nonumber\\
  & 
\qquad+y^{2-4\epsilon}\left(132\zeta_{3}\zeta_{2}-720\zeta_{2}-308\zeta_{3}+315\zeta_{4}-105\zeta_{5}+1008\right)\bigg]
\nonumber\\ &
+ {\cal O}(\epsilon^3) + {\cal O}(y^3)\,,
\end{align}
where terms up to ${\cal O}(\epsilon^6)$ have been computed. For
brevity only terms up to order $\epsilon^2$ are shown.

We have used the described algorithm for all needed 
three-particle initial conditions with one exception: the result
of the integral  $\text{BT9}(1,0,1,0,1,1,1,0,1,1,0,0)$
where the lines $\{1,7,9\}$ are cut is taken over from 
Eq.~(5.32) of Ref.~\cite{Anastasiou:2015yha}.

As a cross check we have computed more integrals in the soft limit than
actually necessary to fix the boundary conditions. Afterwards we have checked
that the solution of the differential equation reproduces these additional
terms.

Note that the algorithm described in this section can also
be applied to the four-particle-cut contribution after applying obvious
modifications. In this way we have cross checked most of
our results, which have been obtained using the method which we
describe in the next section.

%- }}}
%- {{{ 4-particle cuts

\section{\label{sec::4p}Four-particle cuts}

To compute the initial condition of the four-particle-cut contributions we
closely follow the procedure described in Ref.~\cite{Anastasiou:2013srw}.  For
completeness we briefly repeat the individual steps in this section.  The soft
expansion of the four-particle cut integrals exhibit only one region, which is
defined by the scaling in $y$ of the scalar products $s_{ij}$ defined in
Eq.~(\ref{eq:sij}).  Reversed unitarity~\cite{Anastasiou:2002yz} allows for an
expansion in the limit $y\to0$ of the Higgs boson propagator which in our
parametrization is given by
\begin{align}
  &y\left(\frac{1}{(p_1+p_2- p_3-p_4-p_5)^2-x}\right)_c\nonumber\\
  &=\sum_{k=0}^{\infty}\frac{y^{k}}{s}\left[-(s_{34}+s_{35}+s_{45})\right]^k
  \left(\frac{1}{\left(1+s_{13}+s_{23}+s_{14}+s_{24}+s_{15}+s_{25}\right)^{k+1}}
  \right)_c\,,
\end{align}
where the subscript ``$c$'' reminds that the propagator has to be cut.
In the soft limit only the term $k=0$ is needed.  The massless propagators of
the quarks and gluons are expanded as a Taylor series in the limit $y\to0$ as
well. This yields shifts in indices of the propagators, which are removed by
subsequently applying the Laporta algorithm~\cite{Laporta:2001dd} as
implemented in {\tt FIRE}~\cite{Smirnov:2014hma} in the soft kinematics.  We
obtain eleven master integrals. Ten are given in
Ref.~\cite{Anastasiou:2013srw} where analytical results are derived.  The
eleventh integral corresponds to the soft limit of
$\text{BT1}(1,1,1,0,0,0,1,1,0,1,0,0)$ (cf. Fig.~\ref{fig::int_fam}) which can
be cast in the form
\begin{eqnarray}
  F_{11}(\epsilon) &=& \frac{1}{ \Phi_{4}^{s} }
  \int 
  \frac{{\rm d} \Phi_{4}^{s} }{(s_{13}+s_{14})(s_{14}+s_{15})}
  \,,
  \label{eq::F11}
\end{eqnarray}
where $\Phi_{4}^{s}$ is defined in analogy to $\Phi_{3}^{s}$ in Eq.~(\ref{eq::phi3}).
In Ref.~\cite{Anastasiou:2013srw} this integral probably only contributes to
higher orders in $y$ which is why it has not been discussed in that paper.

Following Ref.~\cite{Anastasiou:2013srw} we apply
Eq.~(\ref{eq:repDenomMB}) to convert the sums
in the denominator of Eq.~(\ref{eq::F11})
into products at the cost of introducing Mellin-Barnes integrals.

Introducing energies and angles in analogy to
Eqs.~(\ref{eq:sij}) and~(\ref{eq:pi}) one can integrate the 
energies in terms of $\Gamma$
functions, such that the only non-trivial integrations are given by
three integrations over solid-angles, each of the form of
Eq.~(\ref{eq::omega_int}), which are turned into Mellin-Barnes integrals.
Following this procedure, we arrive at a one-dimensional Mellin-Barnes
integral
\begin{eqnarray}
  F_{11}(\epsilon) &=& 
  \int_{-i\infty}^{+i\infty}\frac{{\rm d}z}{2\pi i}
  \frac{\Gamma(6 - 6 \epsilon) \Gamma(-2 \epsilon - z) 
    \Gamma(-\epsilon - z) \Gamma(-z) 
    \Gamma(1 + z) \Gamma(1 - \epsilon + z)}{\Gamma(4 - 6 \epsilon) 
    \Gamma^2(1 - \epsilon) \Gamma(1 - 2 \epsilon - z)}
  \,,
\end{eqnarray}
which we expand in $\epsilon$ and solve by applying the algorithm 
of Ref.~\cite{Anzai:2012xw}. As a final result we obtain
\begin{eqnarray}
  \label{eq:anaF11}
  F_{11}(\epsilon) &=& 20 \zeta_2+\epsilon (-54 \zeta_2+140
  \zeta_3)+\epsilon^2 (36 \zeta_2-378 \zeta_3+600
  \zeta_4)
  \nonumber\\ &&\mbox{}+\epsilon^3 (252 \zeta_3+160 \zeta_2
  \zeta_3-1620 \zeta_4+1860 \zeta_5)
  \nonumber\\ &&\mbox{}+\epsilon^4 (-432
  \zeta_2 \zeta_3+560 \zeta_3^2+1080 \zeta_4-5022 \zeta_5+6420
  \zeta_6)
  \nonumber\\ &&\mbox{}+\epsilon^5 (288 \zeta_2 \zeta_3-1512
  \zeta_3^2+4800 \zeta_3 \zeta_4+3348 \zeta_5+960 \zeta_2
  \zeta_5-17334 \zeta_6+15240
  \zeta_7)
  \nonumber\\ &&\mbox{}+\mathcal{O}(\epsilon^6)
  \,,
\end{eqnarray}
which we have checked numerically using the package {\tt
  MB.m}~\cite{Czakon:2005rk}. We have also rederived the
integrals\footnote{The integral $F_1(\epsilon)$ is simply the volume of
  four-particle phase space itself.}
$F_2(\epsilon), \dots, F_{10}(\epsilon)$ of
Ref.~\cite{Anastasiou:2013srw}. It is interesting to note, that all
coefficients of Eq.~(\ref{eq:anaF11}) are integers, an observation
also made in Ref.~\cite{Anastasiou:2013srw} for the integrals
$F_2(\epsilon), \dots, F_{10}(\epsilon)$.

For many master integrals, we computed more terms in the soft expansion than
required to fix the integration constants.  These terms could be compared to
the expansion of the exact result and thus strong consistency checks are
obtained.

Note that an alternative method to compute four-particle phase-space
integrals in the soft limit has been developed in Ref.~\cite{Zhu:2014fma}.

%- }}}
%- {{{ Iterated integrals

\section{\label{sec::iter}Iterated integrals beyond HPLs}

\newcommand{\NNNLOqpBTthree}{\text{BT3}}

\begin{figure}[htpbs]
  \centering
  \includegraphics[width=.3\textwidth]{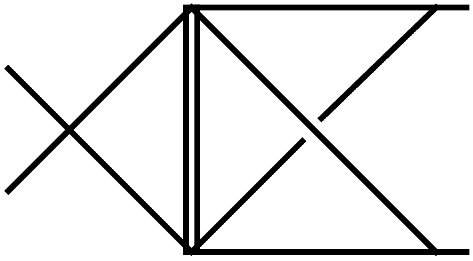}
  \caption{Common subtopology of all the graphs in BT3 which
    generate square root letters.}
  \label{fig:sqrtSubtopo}
\end{figure}

The solution of 16 out of our 17 families can be expressed in terms of
HPLs~\cite{Remiddi:1999ew}, however, for {\rm BT3} this is not possible.
In fact, the
differential equation of the canonical basis implies an alphabet for
the iterated integrals which involves square roots. The letters are
\begin{align}
  \label{eq:sqrtAlphabet}
  \left\{
    \frac{1}{x},\frac{1}{1-x},\frac{1}{1+x},\frac{1}{1+4x},\frac{1}{x\sqrt{1+4x}}
  \right\}
  \,.
\end{align}
The master integrals in which the last two letters show up can be classified
as having the common subtopology drawn in Fig.~\ref{fig:sqrtSubtopo}.  The
contributing integrals with this property are
\begin{align}
  &\NNNLOqpBTthree(1, 0, 0, 1, 1, 1, 1, 1, 1, 0, 0, 0),\nonumber\\
  &\NNNLOqpBTthree(1, 0, -1, 1, 1, 1, 1, 1, 1, 0, 0, 0),\nonumber\\
  &\NNNLOqpBTthree(1, 0, 0, 1, 1, 1, 1, 1, 1, 1, 0, 0),\nonumber\\
  &\NNNLOqpBTthree(1, 1, 1, 1, 1, 1, 1, 1, 1, 0, 0, 0),\nonumber\\
  &\NNNLOqpBTthree(1, 1, 1, 1, 1, 1, 1, 1, 1, 0, -1, 0),\nonumber\\
  &\NNNLOqpBTthree(1, 1, 1, 1, 1, 1, 1, 1, 1, -1, 0, 0)\,.
  \label{eq::BT3}
\end{align}

In general the occurrence of square roots $\sqrt{x-x_0}$ can be
anticipated by observing half-integers in the diagonalized form of the
matrix residue at $x_0$, as shall be briefly explained in the following
using the above example.  Let us denote the system of differential
equations for the integrals in Eq.~(\ref{eq::BT3}) by
$\partial_x \tilde{f} = \tilde{A} \tilde{f}$.  
We expand the matrix elements of $\tilde{A}$ in
a Laurent series around $x_0=-1/4$ and take the coefficient
of $(x-x_0)^{-1}$, which is called the matrix residue. After 
diagonalization we obtain
\begin{align}
   \text{diag}(0,0,0,0,0,1/2-4\ep)\,.
\end{align}
In a next step we expand the element $\tilde{f}_6$ corresponding to the last
entry of the diagonal matrix in a power series in $(x-x_0)$ and obtain with
the help of the differential equation $\partial_x \tilde{f}_6 = (1/2-4\ep)
\tilde{f}_6 / (x-x_0)$.  Note that the occurrence of the half-integer
prefactor on the right-hand side (for $\ep\rightarrow 0$) implies the
occurrence of the square root $\sqrt{1+4x}$, which in the full solution may
show up in coefficients and in the alphabet of iterated integrals.

The present calculation contrasts earlier ones encountering square root
letters~\cite{Aglietti:2004tq,Ablinger:2014uka,Ablinger:2014vwa}, 
where the occurrence of the square root is connected
with the presence of massive two-particle or four-particle cuts in the integrals
(cf. the connection of square root letters in iterated integrals with
(inverse) binomial sums in~\cite{Ablinger:2014bra}, as well as
calculations of Feynman diagrams involving (inverse) binomial sums
in~\cite{Kalmykov:2000qe,Jegerlehner:2002em,Davydychev:2000na,Fleischer:1998nb}). 
Topology {\rm BT3}, however, represents four-particle 
phase-space integrals with only one massive line in the final state.

The canonical differential equation can be solved, as usual,
order-by-order in $\epsilon$. Afterwards the constants of integration have
to be determined. This is done by expanding the generic solution in a
generalized Taylor series expansion around $x=1$ and matching with a
calculation in the soft limit. For the expansion one needs to extract the
logarithmic part due to $\log(1-x)$. This can be done using the shuffle
algebra, and making sure that $1/(1-x)$ never occurs in the rightmost index of
the iterated integrals.  As a result the iterated integrals either diverge
like $\log(1-x)$ or are regular in the limit $x\rightarrow 1$. For the
matching procedure one now only needs the $(1-x)^0$-order, i.e. the regular
part evaluated at $x=1$, while logarithmic orders provide a cross check for
the generic solution with the calculation of the boundary conditions.

In this way the canonical master integrals and hence the Laporta masters are
expressed in terms of iterated integrals over the
alphabet~(\ref{eq:sqrtAlphabet}). For numerical evaluations, it is 
advantageous to modify the above alphabet to be
\begin{align}
  \left\{
    \frac{1}{x}, \frac{1}{1-x}, \frac{1}{1+x}, \frac{1}{1+4x},
    \frac{1}{x}\left(\frac{1}{\sqrt{1+4x}}-1\right) 
  \right\}\,,
\end{align}
so only one letter is singular as $x \rightarrow 0$. 

The contributions to the single Higgs boson production amplitude do
not span the full space of functions generated by the above
alphabet. In fact the relevant iterated integrals involving the
square root letter can be constructed from
\begin{align}
  f_0 = \frac{1}{x}\,, \qquad
  f_{-1} =  \frac{1}{1+x}\,, \qquad
  f_{s4} = \frac{1}{x}\left(\frac{1}{\sqrt{1+4x}}-1\right) \,.
  \label{eq::fs4}
\end{align}

For the treatment of algebraic relations and the series expansions of
the iterated integrals with square root letters, the package
{\tt HarmonicSums} was
used~\cite{Ablinger:2010kw,Ablinger:2013hcp,Ablinger:2014wca}.  For the
numerical implementation, the convergent 
series expansions around $x=0$ and $x=1$ are helpful, which are
available once the letter $1/x$ is shuffled away from the rightmost
position in the indices of the iterated integrals. Unfortunately in
contrast to the case of HPLs~\cite{Remiddi:1999ew,Gehrmann:2001pz},
the series expansion around $x=0$ has a radius of convergence of $1/4$,
thus more terms in the expansions are needed.

The iterated integrals involving square root letters were implemented
numerically in {\tt Mathematica}, using series expansions for functions of
weight $\le$3 and up to twofold numerical integrals. In this way we are
able to yield 10 good digits at the timescale of a second and below for
the most complicated functions at weight 5.

%- }}}
%- {{{ Results

\section{\label{sec::res}Results}

The total partonic cross section can be written as
\begin{eqnarray}
  \hat\sigma &=& C_1^2 \tilde\sigma\,,
\end{eqnarray}
where $C_1^2$ and $\tilde\sigma$ are separately finite
after renormalization~\cite{Chetyrkin:1997un}
and the convolution of the lower-order cross sections with the
splitting functions~\cite{Hoschele:2012xc,Hoeschele:2013gga,Buehler:2013fha}. 

Our final result can be cast in the form
\begin{eqnarray}
  \tilde{\sigma}(qq^\prime\to H+X) &=& 
  A\left[
  \left( \frac{\alpha_s}{\pi}\right)^2
  \tilde{\sigma}_{qq^\prime}^{(2)}
  +\left( \frac{\alpha_s}{\pi}\right)^3
  \tilde{\sigma}_{qq^\prime}^{(3)}
  \right]
  \,,
\end{eqnarray}
where $A = G_F \pi /(32\sqrt{2})$
and $\tilde{\sigma}_{qq^\prime}^{(2)}$ is given in Eq.~(54) of
Ref.~\cite{Anastasiou:2002yz} and $\tilde{\sigma}_{qq^\prime}^{(3)}$ reads
after identifying renormalization and factorization scale with the Higgs boson
mass (i.e. $\mu_r=\mu_f=m_h$) 
\begin{align}
\tilde{\sigma}_{qq^\prime}^{(3)} ={}&
 n_l \Biggl[
     \frac{1}{27} (x+2)^2 H_0^4(x)+\frac{2}{243} \left(11 x^2-40 x+188\right) H_0^3(x)-\frac{4}{27} (x+2)^2 \zeta_2 H_0^2(x)
\N\\&
    -\frac{1}{243} \left(772 x^2+1156 x-2213\right) H_0^2(x)-\frac{16}{9} (x-1) (x+3) H_1(x) H_0^2(x)
\N\\&
    -\frac{1}{243} \left(3139 x^2+5218 x-7620\right) H_0(x) -\frac{40}{27} (x-1) (x+3) H_1^2(x) H_0(x)
\N\\&
   +\frac{2}{27} \left(83 x^2+264 x+148\right) H_{0,1,1}(x)+\frac{4}{81} \left(107 x^2+344 x+164\right) H_{0,1}(x) H_0(x)
\N\\&
   -\frac{32}{27} \left(x^2+x+4\right) \zeta_2 H_0(x)-\frac{70}{81} (x-1) (13 x+33) H_1(x) H_0(x)
\N\\&
   -\frac{56}{27} (x+2)^2 H_{0,0,1}(x) H_0(x)+\frac{40}{27} (x+2)^2 H_{0,1,1}(x) H_0(x)-\frac{98}{135} (x+2)^2 \zeta_2^2
\N\\&
   -\frac{4}{27} (x+2)^2 \zeta_{3} H_0(x)-\frac{8}{27} (x-1) (x+3) H_1^3(x)-\frac{1}{324} (x-1) (2549 x+15343)
\N\\&
   -\frac{1}{54} (x-1) (319 x+771) H_1^2(x)-\frac{11}{27} (x+2)^2 H_{0,1}^2(x)-\frac{8}{27} (x+2)^2 \zeta_2 H_{0,1}(x)
\N\\&
   +\frac{2}{243} \left(215 x^2-904 x-1402\right) \zeta_2+\frac{2}{243} \left(469 x^2+1840 x-218\right) H_{0,1}(x)
\N\\&
   -\frac{41}{486} (x-1) (169 x+519) H_1(x)-\frac{4}{81} \left(119 x^2+464 x+260\right) H_{0,0,1}(x)
\N\\&
   +\frac{56}{27} (x+2)^2 H_{0,0,0,1}(x)+\frac{22}{27} (x+2)^2 H_{0,1,0,1}(x)+\frac{16}{27} (x-1) (x+3) \zeta_2 H_1(x)
\N\\&
   +\frac{8}{9} (x+2)^2 H_{0,1}(x) H_0^2(x)+\frac{8}{9} (x+2)^2 H_{0,1,1,1}(x)+\frac{2}{81} \left(13 x^2+184 x+4\right) \zeta_{3}\Biggr]
\N\\&
   +\frac{1}{729} \left(1064 x^3-2853 x^2+107433 x-41149\right) H_0^3(x)-\frac{112}{27} (x+2)^2 H_{0,1}(x) H_0^3(x)
\N\\&
   -\frac{1}{27} \left(13 x^2+135 x+164\right) H_0^4(x)+\frac{4}{27} \left(29 x^2-28 x+28\right) \zeta_2 H_0^3(x)
\N\\&
   +\frac{8}{81} (x+1) (97 x-294) H_{-1}(x) H_0^3(x)+\frac{7}{81} (x-1) (97 x+285) H_1(x) H_0^3(x)
\N\\&
   -\frac{1}{810} \left(175 x^2-308 x+216\right) H_0^5(x)-\frac{130}{27} (x-2)^2 H_{0,-1}(x) H_0^3(x)
\N\\&
   -\frac{356}{27} (x-3) (x+1) H_{-1}^2(x) H_0^2(x)+\frac{836}{27} (x-1) (x+3) H_1^2(x) H_0^2(x)
\N\\&
   +\frac{1}{5832} \left(46480 x^3-286656 x^2+3753336 x-1756017\right) H_0^2(x)
\N\\&
   +\frac{35}{216}(118 x+85) \sqrt{4 x+1} H_0^2(x)-\frac{2}{81} \left(105 x^2-1548 x-1900\right) \zeta_2 H_0^2(x)
\N\\&
   -\frac{35}{54} \left(9 x^2-2 x-8\right) H_{s4}(x) H_0^2(x)+\frac{70}{9} (x-3) (x+1) H_{-1,s4}(x) H_0^2(x)
\N\\&
   -\frac{35}{9} (x-6) x H_{0,s4}(1) H_0^2(x)-\frac{1}{81} \left(1083 x^2-2556 x-3046\right) H_{0,-1}(x) H_0^2(x)
\N\\&
   +\frac{1}{486} (x+1) \left(80 x^2+12889 x-46117\right) H_{-1}(x) H_0^2(x)
\N\\&
   -\frac{35}{9} (x-2)^2 H_{0,-1,s4}(x) H_0^2(x)+\frac{1}{54} (x-1) \left(400 x^2-1019 x+35945\right) H_1(x) H_0^2(x)
\N\\&
   -\frac{35}{9} (x-6) x H_{0,s4}(x) H_0^2(x)+\frac{1}{27} \left(1220 x^2-4844 x-7485\right) H_{0,1}(x) H_0^2(x)
\N\\&
   +\frac{356}{27} (x-2)^2 H_{0,-1,-1}(x) H_0^2(x)+\frac{140}{9} (x-3) (x+1) H_{0,s4}(1) H_{-1}(x) H_0(x)
\N\\&
   -\frac{2}{27} \left(317 x^2-1612 x+212\right) H_{0,0,1}(x) H_0^2(x)-\frac{836}{27} (x+2)^2 H_{0,1,1}(x) H_0^2(x)
\N\\&
   +\frac{8}{9} \left(13 x^2+31 x+30\right) \zeta_{3} H_0^2(x)+\frac{592}{81} (x-3) (x+1) H_{-1}^3(x) H_0(x)
\N\\&
   +\frac{3212}{81} (x-1) (x+3) H_1^3(x) H_0(x)-\frac{4}{135} \left(361 x^2-1044 x+36\right) \zeta_2^2 H_0(x)
\N\\&
   +\frac{28}{81} \log^4(2) (x-2)^2 H_0(x)-\frac{2}{243} (x+1) \left(104 x^2+1633 x-7915\right) H_{-1}^2(x) H_0(x)
\N\\&
   +\frac{224}{27} \Li_4\left(\frac{1}{2}\right) (x-2)^2 H_0(x)+\frac{14}{729} (x+1) \left(280 x^2-1213 x-5327\right) H_{-1}(x) H_0(x)
\N\\&
   +\frac{1}{54} (x-1) \left(592 x^2-847 x+53333\right) H_1^2(x) H_0(x)-\frac{160}{27} (x-2)^2 H_{0,-1}^2(x) H_0(x)
\N\\&
   -\frac{1}{8748} \left(421040 x^3-394707 x^2-12461502 x+7407221\right) H_0(x)
\N\\&
   -\frac{1}{9} \left(79 x^2-1028 x-212\right) H_{0,1}^2(x) H_0(x)-\frac{56}{27} \log^2(2) (x-2)^2 \zeta_2 H_0(x)
\N\\&
   -\frac{1}{243} \left(3008 x^3-10503 x^2+202176 x-46836\right) \zeta_2 H_0(x)
\N\\&
   +\frac{35}{108} \sqrt{4 x+1} (118 x+85) H_{s4}(x) H_0(x)-\frac{35}{27} \left(9 x^2-2 x-8\right) H_{0,s4}(1) H_0(x)
\N\\&
   -\frac{35}{9} (x-2)^2 \zeta_2 H_{0,s4}(1) H_0(x)+\frac{70}{9} (x-2)^2 H_{-1,s4}(1) H_{0,s4}(1) H_0(x)
\N\\&
   +\frac{140}{9} (x-3) (x+1) H_{-1,s4,s4}(x) H_0(x)-\frac{140}{9} (x-3) (x+1) H_{0,-1,s4}(x) H_0(x)
\N\\&
   +\frac{70}{9} \log(2) (x-2)^2 H_{0,0,s4}(1) H_0(x)-\frac{70}{9} (x-6) x H_{0,0,s4}(1) H_0(x)
\N\\&
   +\frac{70}{9} (x-6) x H_{0,0,s4}(x) H_0(x)-\frac{70}{9} \log(2) (x-2)^2 H_{0,s4,s4}(1) H_0(x)
\N\\&
   +\frac{70}{9} (x-6) x H_{0,s4,s4}(1) H_0(x)-\frac{70}{9} (x-6) x H_{0,s4,s4}(x) H_0(x)
\N\\&
   -\frac{70}{9} (x-2)^2 H_{-1,0,0,s4}(1) H_0(x)-\frac{70}{9} (x-2)^2 H_{-1,0,s4,s4}(1) H_0(x)
\N\\&
   -\frac{70}{9} (x-2)^2 H_{-1,s4,0,s4}(1) H_0(x)-\frac{70}{9} (x-2)^2 H_{0,-1,s4,s4}(1) H_0(x)
\N\\&
   -\frac{70}{9} (x-2)^2 H_{0,-1,s4,s4}(x) H_0(x)+\frac{70}{9} (x-2)^2 H_{0,0,-1,s4}(1) H_0(x)
\N\\&
   +\frac{200}{9} (x-2)^2 H_{0,0,-1}(x) H_0^2(x)-\frac{1}{81} (x-1) \left(908 x^2+7916 x-135025\right) H_1(x) H_0(x)
\N\\&
   -\frac{35}{54} \left(9 x^2-2 x-8\right) H_{s4}(x)^2 H_0(x)-\frac{4}{27} (x+1) (373 x-1101) \zeta_2 H_{-1}(x) H_0(x)
\N\\&
   -\frac{8}{81} \left(306 x^2-474 x-497\right) H_{0,-1,-1}(x) H_0(x)-\frac{4}{27} (x-1) (307 x+963) \zeta_2 H_1(x) H_0(x)
\N\\&
   +\frac{1}{243} \left(16 x^3-5013 x^2+4140 x+28873\right) H_{0,-1}(x) H_0(x)-\frac{35}{9} (x-2)^2 \zeta_2 H_{0,s4,s4}(1)
\N\\&
   +\frac{740}{27} (x-2)^2 \zeta_2 H_{0,-1}(x) H_0(x)+\frac{8}{9} (x+1) (27 x-79) H_{-1}(x) H_{0,-1}(x) H_0(x)
\N\\&
   -\frac{70}{9} (x-2)^2 H_{0,s4}(1) H_{0,-1}(x) H_0(x)-\frac{8}{27} (x-1) (25 x+117) H_1(x) H_{0,-1}(x) H_0(x)
\N\\&
   +\frac{1}{243} \left(2960 x^3+18009 x^2-2034 x+69166\right) H_{0,1}(x) H_0(x)
\N\\&
   -\frac{448}{27} (x-2)^2 H_{0,-1}(x) H_{0,1}(x) H_0(x)+\frac{8}{9} (x+1) (37 x-113) H_{-1}(x) H_{0,1}(x) H_0(x)
\N\\&
   +\frac{628}{27} (x+2)^2 \zeta_2 H_{0,1}(x) H_0(x)+\frac{2}{27} (x-1) (475 x+1383) H_1(x) H_{0,1}(x) H_0(x)
\N\\&
   +\frac{140}{9} (x-2)^2 H_{0,0,-1,s4}(x) H_0(x)-\frac{16}{27} \left(43 x^2-160 x-111\right) H_{0,-1,1}(x) H_0(x)
\N\\&
   -\frac{496}{27} (x-2)^2 H_{0,0,-1,-1}(x) H_0(x)+\frac{4}{27} \left(81 x^2-620 x-112\right) H_{0,0,-1}(x) H_0(x)
\N\\&
   +\frac{896}{27} (x-2)^2 H_{0,0,-1,1}(x) H_0(x)-\frac{2}{81} \left(3996 x^2-17088 x-31681\right) H_{0,0,1}(x) H_0(x)
\N\\&
   -52 (x-2)^2 H_{0,0,0,-1}(x) H_0(x)-\frac{16}{27} \left(43 x^2-160 x-111\right) H_{0,1,-1}(x) H_0(x)
\N\\&
   +\frac{2}{27} \left(923 x^2-8188 x-6336\right) H_{0,1,1}(x) H_0(x)-\frac{592}{27} (x-2)^2 H_{0,-1,-1,-1}(x) H_0(x)
\N\\&
   +\frac{4}{9} \left(275 x^2-820 x+484\right) H_{0,0,0,1}(x) H_0(x)+\frac{896}{27} (x-2)^2 H_{0,0,1,-1}(x) H_0(x)
\N\\&
   +\frac{64}{27} \left(9 x^2-20 x+36\right) H_{0,1,0,-1}(x) H_0(x)+\frac{4}{27} \left(37 x^2-838 x+842\right) \zeta_{3} H_0(x)
\N\\&
   +\frac{2}{9} \left(x^2-1340 x-524\right) H_{0,1,0,1}(x) H_0(x)-\frac{3212}{27} (x+2)^2 H_{0,1,1,1}(x) H_0(x)
\N\\&
   +\frac{196}{27} \log(2) (x-2)^2 \zeta_{3} H_0(x)-\frac{500}{9} (x-1) (x+3) \zeta_2 H_1^2(x)
\N\\&
   +\frac{1}{54} (x-1) \left(256 x^2-237 x+23559\right) H_1^3(x)+\frac{2}{135} \left(2269 x^2-1338 x+9038\right) \zeta_2^2
\N\\&
   +\frac{35}{9} (x-6) x H_{0,s4}(1)^2+\frac{35}{9} (x-6) x H_{0,s4}(x)^2+\frac{872}{27} (x-3) (x+1) \zeta_2 H_{-1}^2(x)
\N\\&
   +\frac{1208}{81} (x-1) (x+3) H_1^4(x)-\frac{1}{2916} (x-1) \left(24448 x^2+200029 x-3186149\right) H_1^2(x)
\N\\&
   +\frac{4}{81} \left(63 x^2-6 x+214\right) H_{0,-1}^2(x)-\frac{1}{27} \left(494 x^2-5212 x-4845\right) H_{0,1}^2(x)
\N\\&
   +\frac{1}{26244} (x-1) \left(57136 x^2-1139639 x+29021665\right)+\frac{112}{27} \log^2(2) (x-3) (x+1) \zeta_2
\N\\&
   -\frac{1}{1458} \left(8032 x^3-239799 x^2+1426872 x-432084\right) \zeta_2-\frac{56}{81} \log^4(2) (x-3) (x+1)
\N\\&
   -\frac{448}{27} \Li_4\left(\frac{1}{2}\right) (x-3) (x+1)+\frac{35}{108} \sqrt{4 x+1} (118 x+85) H_{0,s4}(1)
\N\\&
   +\frac{70}{9} (x-3) (x+1) \zeta_2 H_{0,s4}(1)-\frac{35}{27} \left(9 x^2-2 x-8\right) H_{s4}(x) H_{0,s4}(1)
\N\\&
   -\frac{140}{9} (x-3) (x+1) H_{-1,s4}(1) H_{0,s4}(1)+\frac{140}{9} (x-3) (x+1) H_{-1,s4}(x) H_{0,s4}(1)
\N\\&
   -\frac{35}{108} \sqrt{4 x+1} (118 x+85) H_{0,s4}(x)+\frac{35}{27} \left(9 x^2-2 x-8\right) H_{s4}(x) H_{0,s4}(x)
\N\\&
   -\frac{70}{9} (x-6) x H_{0,s4}(1) H_{0,s4}(x)+\frac{70}{9} (x-2)^2 H_{0,s4}(1) H_{0,-1,s4}(1)
\N\\&
   -\frac{70}{9} (x-2)^2 H_{0,s4}(1) H_{0,-1,s4}(x)-\frac{140}{9} \log(2) (x-3) (x+1) H_{0,0,s4}(1)
\N\\&
   -\frac{35}{27} \left(9 x^2-2 x-8\right) H_{0,0,s4}(1)+\frac{35}{27} \left(9 x^2-2 x-8\right) H_{0,0,s4}(x)
\N\\&
   +\frac{140}{9} \log(2) (x-3) (x+1) H_{0,s4,s4}(1)+\frac{35}{27} \left(9 x^2-2 x-8\right) H_{0,s4,s4}(1)
\N\\&
   -\frac{35}{27} \left(9 x^2-2 x-8\right) H_{0,s4,s4}(x) +\frac{140}{9} (x-3) (x+1) H_{-1,0,0,s4}(1)
\N\\&
   -\frac{140}{9} (x-3) (x+1) H_{-1,0,0,s4}(x)+\frac{140}{9} (x-3) (x+1) H_{-1,0,s4,s4}(1)
\N\\&
   -\frac{140}{9} (x-3) (x+1) H_{-1,0,s4,s4}(x)+\frac{140}{9} (x-3) (x+1) H_{-1,s4,0,s4}(1)
\N\\&
   -\frac{140}{9} (x-3) (x+1) H_{-1,s4,0,s4}(x)+\frac{140}{9} (x-3) (x+1) H_{0,-1,s4,s4}(1)
\N\\&
   -\frac{140}{9} (x-3) (x+1) H_{0,-1,s4,s4}(x)-\frac{140}{9} (x-3) (x+1) H_{0,0,-1,s4}(1)
\N\\&
   +\frac{140}{9} (x-3) (x+1) H_{0,0,-1,s4}(x)-\frac{70}{9} (x-2)^2 H_{0,-1,0,s4,s4}(1)
\N\\&
   +\frac{70}{9} (x-2)^2 H_{0,-1,0,s4,s4}(x)-\frac{70}{9} (x-2)^2 H_{0,-1,s4,0,s4}(1)
\N\\&
   +\frac{70}{9} (x-2)^2 H_{0,-1,s4,0,s4}(x)+\frac{140}{9} (x-2)^2 H_{0,0,-1,0,s4}(1)
\N\\&
   -\frac{140}{9} (x-2)^2 H_{0,0,-1,0,s4}(x)-\frac{140}{9} (x-2)^2 H_{0,0,-1,s4,s4}(1)
\N\\&
   +\frac{140}{9} (x-2)^2 H_{0,0,-1,s4,s4}(x)+\frac{140}{3} (x-2)^2 H_{0,0,0,-1,s4}(1)
\N\\&
   -\frac{140}{3} (x-2)^2 H_{0,0,0,-1,s4}(x)+\frac{70}{3} (x-2)^2 H_{0,0,0,s4,-1}(1)
\N\\&
   -\frac{70}{3} (x-2)^2 H_{0,0,0,s4,-1}(x)+\frac{70}{9} (x-2)^2 H_{0,0,s4,0,-1}(1)
\N\\&
   -\frac{70}{9} (x-2)^2 H_{0,0,s4,0,-1}(x)-\frac{4}{27} (x+1) \left(4 x^2+400 x-1489\right) \zeta_2 H_{-1}(x)
\N\\&
   +\frac{140}{9} (x-3) (x+1) H_{0,0,s4}(1) H_{-1}(x)-\frac{140}{9} (x-3) (x+1) H_{0,s4,s4}(1) H_{-1}(x)
\N\\&
   -\frac{1}{1458} (x-1) \left(52112 x^2-42806 x-2393137\right) H_1(x)+\frac{64}{3} (x-2)^2 H_{0,1}(x) H_{0,1,-1}(x)
\N\\&
   -\frac{8}{243} (x-1) \left(418 x^2-2777 x+32665\right) \zeta_2 H_1(x)+\frac{70}{9} (x-2)^2 H_{0,0,s4}(x) H_{0,-1}(x)
\N\\&
   +\frac{8}{27} \left(50 x^2-131 x-165\right) \zeta_2 H_{0,-1}(x)-\frac{592}{27} (x-3) (x+1) H_{-1}^2(x) H_{0,-1}(x)
\N\\&
   -\frac{14}{729} (x+1) \left(280 x^2-1213 x-5327\right) H_{0,-1}(x) -\frac{70}{9} (x-2)^2 H_{0,0,s4}(1) H_{0,-1}(x)
\N\\&
   -\frac{140}{9} (x-3) (x+1) H_{0,s4}(1) H_{0,-1}(x) +\frac{70}{9} (x-2)^2 H_{0,s4,s4}(1) H_{0,-1}(x)
\N\\&
   +\frac{4}{243} (x+1) \left(104 x^2+1633 x-7915\right) H_{-1}(x) H_{0,-1}(x)
\N\\&
   -\frac{64}{3} (x-3) (x+1) H_{-1}^2(x) H_{0,1}(x)+\frac{68}{27} (x-1) (x+3) H_1^2(x) H_{0,1}(x)
\N\\&
   +\frac{1}{1458} \left(21656 x^3-106251 x^2-38232 x+710568\right) H_{0,1}(x)
\N\\&
   -\frac{4}{81} \left(2373 x^2-4638 x-8990\right) \zeta_2 H_{0,1}(x)+\frac{256}{27} (x-2)^2 H_{0,1}(x) H_{0,0,-1}(x)
\N\\&
   -\frac{2}{243} (x+1) \left(32 x^2-5567 x+18887\right) H_{-1}(x) H_{0,1}(x)
\N\\&
   -\frac{8}{243} (x-1) \left(14 x^2+194 x-355\right) H_1(x) H_{0,1}(x)+\frac{4}{81} \left(51 x^2-430\right) H_{0,-1}(x) H_{0,1}(x)
\N\\&
   -\frac{4}{243} (x+1) \left(104 x^2+1633 x-7915\right) H_{0,-1,-1}(x)-\frac{872}{27} (x-2)^2 \zeta_2 H_{0,-1,-1}(x)
\N\\&
   +\frac{1184}{27} (x-3) (x+1) H_{-1}(x) H_{0,-1,-1}(x)+\frac{592}{27} (x-2)^2 H_{0,-1}(x) H_{0,-1,-1}(x)
\N\\&
   +\frac{64}{3} (x-2)^2 H_{0,1}(x) H_{0,-1,-1}(x)+\frac{2}{243} (x+1) \left(32 x^2-5567 x+18887\right) H_{0,-1,1}(x)
\N\\&
   +\frac{128}{3} (x-3) (x+1) H_{-1}(x) H_{0,-1,1}(x)-\frac{32}{3} (x-2)^2 H_{0,-1}(x) H_{0,-1,1}(x)
\N\\&
   -\frac{1}{243} \left(112 x^3+2943 x^2-24948 x+11629\right) H_{0,0,-1}(x)-\frac{1048}{27} (x-2)^2 \zeta_2 H_{0,0,-1}(x)
\N\\&
   +\frac{140}{9} (x-2)^2 H_{0,s4}(1) H_{0,0,-1}(x)+\frac{32}{27} (x+1) (4 x-15) H_{-1}(x) H_{0,0,-1}(x)
\N\\&
   +\frac{16}{27} (x-1) (25 x+117) H_1(x) H_{0,0,-1}(x)-\frac{8}{3} (x-2)^2 H_{0,-1}(x) H_{0,0,-1}(x)
\N\\&
   -\frac{1}{243} \left(4112 x^3+17235 x^2+8928 x-11415\right) H_{0,0,1}(x)
\N\\&
   +\frac{16}{27} \left(83 x^2-316 x+68\right) \zeta_2 H_{0,0,1}(x)-\frac{16}{9} (x+1) (7 x-23) H_{-1}(x) H_{0,0,1}(x)
\N\\&
   -\frac{4}{27} (x-1) (407 x+1179) H_1(x) H_{0,0,1}(x)+\frac{128}{3} (x-3) (x+1) H_{-1}(x) H_{0,1,-1}(x)
\N\\&
   +\frac{4}{81} \left(529 x^2-2060 x+620\right) H_{0,1}(x) H_{0,0,1}(x)+\frac{176}{27} (x-2)^2 H_{0,-1}(x) H_{0,0,1}(x)
\N\\&
   +\frac{2}{243} (x+1) \left(32 x^2-5567 x+18887\right) H_{0,1,-1}(x)-\frac{32}{3} (x-2)^2 H_{0,-1}(x) H_{0,1,-1}(x)
\N\\&
   +\frac{1}{243} \left(3600 x^3+28827 x^2-50220 x+145505\right) H_{0,1,1}(x)+\frac{500}{9} (x+2)^2 \zeta_2 H_{0,1,1}(x)
\N\\&
   +\frac{128}{3} (x-3) (x+1) H_{-1}(x) H_{0,1,1}(x)+\frac{8}{27} (x-1) (x+3) H_1(x) H_{0,1,1}(x)
\N\\&
   -\frac{64}{3} (x-2)^2 H_{0,-1}(x) H_{0,1,1}(x)+\frac{2}{27} \left(27 x^2+1836 x+812\right) H_{0,1}(x) H_{0,1,1}(x)
\N\\&
   -\frac{1184}{27} (x-3) (x+1) H_{0,-1,-1,-1}(x)-\frac{128}{3} (x-3) (x+1) H_{0,-1,-1,1}(x)
\N\\&
   -\frac{128}{3} (x-3) (x+1) H_{0,-1,1,-1}(x)-\frac{128}{3} (x-3) (x+1) H_{0,-1,1,1}(x)
\N\\&
   -\frac{32}{27} (x+1) (4 x-15) H_{0,0,-1,-1}(x)-\frac{8}{81} \left(75 x^2+840 x-718\right) H_{0,0,-1,1}(x)
\N\\&
   -\frac{2}{27} \left(179 x^2-2740 x+22\right) H_{0,0,0,-1}(x)+\frac{2}{27} \left(555 x^2-5808 x-11551\right) H_{0,0,0,1}(x)
\N\\&
   -\frac{8}{81} \left(75 x^2+840 x-718\right) H_{0,0,1,-1}(x)-\frac{128}{3} (x-3) (x+1) H_{0,1,-1,-1}(x)
\N\\&
   -\frac{128}{3} (x-3) (x+1) H_{0,1,-1,1}(x)-\frac{4}{81} \left(201 x^2+552 x-1132\right) H_{0,1,0,-1}(x)
\N\\&
   +\frac{2}{81} \left(2847 x^2-13308 x-17566\right) H_{0,1,0,1}(x)-\frac{128}{3} (x-3) (x+1) H_{0,1,1,-1}(x)
\N\\&
   +\frac{2}{27} \left(1905 x^2-9008 x-8556\right) H_{0,1,1,1}(x)-\frac{1184}{27} (x-2)^2 H_{0,-1,0,-1,-1}(x)
\N\\&
   +\frac{32}{3} (x-2)^2 H_{0,-1,1,0,-1}(x)-\frac{2368}{27} (x-2)^2 H_{0,0,-1,-1,-1}(x)
\N\\&
   +\frac{568}{27} (x-2)^2 H_{0,0,-1,0,-1}(x)+\frac{128}{3} (x-2)^2 H_{0,0,-1,1,1}(x)
\N\\&
   +\frac{568}{9} (x-2)^2 H_{0,0,0,-1,-1}(x)-48 (x-2)^2 H_{0,0,0,-1,1}(x)
\N\\&
   +\frac{512}{9} (x-2)^2 H_{0,0,0,0,-1}(x)-\frac{8}{27} \left(577 x^2-1580 x+1296\right) H_{0,0,0,0,1}(x)
\N\\&
   -48 (x-2)^2 H_{0,0,0,1,-1}(x)-\frac{16}{27} \left(59 x^2-108 x+236\right) H_{0,0,1,0,-1}(x)
\N\\&
   -\frac{4}{27} \left(295 x^2-2996 x-316\right) H_{0,0,1,0,1}(x)-\frac{128}{3} (x-2)^2 H_{0,0,1,1,-1}(x)
\N\\&
   +\frac{32}{3} (x-2)^2 H_{0,1,-1,0,-1}(x)-\frac{64}{3} (x-2)^2 H_{0,1,-1,0,1}(x)
\N\\&
   -\frac{512}{27} \left(x^2+4\right) H_{0,1,0,0,-1}(x)+\frac{4}{81} \left(71 x^2+4460 x+1780\right) H_{0,1,0,0,1}(x)
\N\\&
   -\frac{64}{3} (x-2)^2 H_{0,1,0,1,-1}(x)-\frac{2}{27} \left(151 x^2+5788 x+2716\right) H_{0,1,0,1,1}(x)
\N\\&
   +\frac{64}{3} (x-2)^2 H_{0,1,1,0,-1}(x)-\frac{2}{27} \left(61 x^2+1972 x+948\right) H_{0,1,1,0,1}(x)
\N\\&
   -\frac{4832}{27} (x+2)^2 H_{0,1,1,1,1}(x)+\frac{16}{27} \left(31 x^2+354 x+58\right) \zeta_{5}
\N\\&
   -\frac{392}{27} \log(2) (x-3) (x+1) \zeta_{3}-\frac{2}{243} \left(848 x^3+819 x^2+102132 x-25745\right) \zeta_{3}
\N\\&
   -\frac{2}{27} \left(147 x^2+2564 x+236\right) \zeta_2 \zeta_{3}-\frac{35}{3} (x-2)^2 H_{0,s4}(1) \zeta_{3}
\N\\&
   -\frac{4}{27} (x+1) (445 x-1329) H_{-1}(x) \zeta_{3}-\frac{32}{27} (x-1) (76 x+249) H_1(x) \zeta_{3}
\N\\&
   +\frac{296}{9} (x-2)^2 H_{0,-1}(x) \zeta_{3}+\frac{424}{9} (x+2)^2 H_{0,1}(x) \zeta_{3}
\,.
\label{eq::res}
\end{align}
A computer-readable version of this equation can be obtained
from~\cite{progdata}.  In Eq.~(\ref{eq::res}) $n_l=5$ denotes the number of
massless quarks and $\zeta_n$ stands for Riemann's zeta function evaluated at
$n$.  $H_{\vec{a}}(x)$ where $\vec{a}$ only has the elements $0$ and $\pm1$
denote HPLs~\cite{Remiddi:1999ew}. In case $\vec{a}$ contains also $s4$ the
corresponding function refers to the iterated integral with square-root
element introduced in Eq.~(\ref{eq::fs4}) of Section~\ref{sec::iter}.  In
Eq.~(\ref{eq::res}) we observe iterated integrals up to weight~5.

Some of the iterated integrals in Eq.~(\ref{eq::res}), which are evaluated for
$x=1$, can be transformed to combinations of Riemann zeta functions. However,
we prefer to leave $H_{\vec{a}}(1)$ since these terms disappear by
construction in case Eq.~(\ref{eq::res}) is evaluated for $x=1$.

The square root letter occurring in the result for the topology \text{BT3} has
already been introduced in Ref.~\cite{Ablinger:2014bra}, where it was named
$f_{w_{14}}$. The corresponding iterated integrals occurred in the context of
the calculation of three-loop contributions to massive operator matrix
elements of Ref.~\cite{Ablinger:2014yaa}. Interestingly, using the
substitution $x\rightarrow (1-x^\prime)/x^{\prime\,2}$, the integrals
involving $f_{s4}$ in Eq.~(\ref{eq::res}) can be brought into the form of
cyclotomic polylogarithms (cf. Ref.~\cite{Ablinger:2011te}) and can thus be
represented as Goncharov polylogarithms~\cite{Goncharov:1998kja} with the
sixth root of unity appearing in the indices, more precisely with the alphabet
$\{1,0,(-1)^{1/3}\}$.  Furthermore, all functions without a letter $(-1)$ can
be reduced to HPLs at the cost of a more complicated argument and an increase
of the number of terms.  In this representation the constants introduced via
matching at $x=1$ are cyclotomic/multiple polylogarithms evaluated at the
reciprocal of the golden ratio $x'=(\sqrt{5}-1)/2$.  Nevertheless, since the
$H_{\ldots s4 \ldots}$ are by construction real and since their numerical
implementation is straightforward we decided not to rewrite the expression in
Eq.\ (\ref{eq::res}).

In Ref.~\cite{Anastasiou:2014lda} the second term in the threshold expansion
for the N$^3$LO corrections to Higgs boson production has been
computed. Furthermore, for all contributing partonic channels the exact
dependence on $x$ is provided for the coefficients of the leading logarithms
in $\log(1-x)$. In Eq.~(\ref{eq::res}) only (some of) the HPLs are
divergent in the limit $x\to1$ since in the iterated integrals involving
$s4$ the letter $1/(1-x)$ is absent. After extracting the logarithmic 
divergencies of the HPLs we find full agreement with the results
given in Eqs.~(2.26) and~(2.27) of~\cite{Anastasiou:2014lda} for the 
coefficients of the $\log^3(1-x)$ and $\log^4(1-x)$ contribution,
respectively.\footnote{We thank Claude Duhr for communications concerning
  this point.}

%- }}}
%- {{{ Conclusions

\section{\label{sec::con}Conclusions}

In this paper we have computed a contribution to the third-order partonic
cross section for Higgs boson production in gluon fusion, namely the
sub-process initiated by two quarks with different flavour.  The numerical
impact of this contribution is small. However, we have obtained analytic
results retaining the exact dependence on the Higgs boson mass and the
partonic center-of-mass energy.  This constitutes a new result since to date
only an expansion around the soft limit has been presented in the literature.
Our findings constitute an important step towards an exact result of all
third-order contributions to the Standard Model Higgs boson production.

In the course of our calculation we have mapped all contributing amplitudes to
17 integral families.  For each family we have constructed a canonical basis
and derived the corresponding system of differential equations. After
evaluating the three- and four-particle cut initial conditions the
differential equations could be solved in terms of HPLs in all 
integral families except one, which required additional letters in the 
alphabet of the iterated integrals.

%- }}}

%- {{{ Ackn.:

\section*{Acknowledgments}

We would like to thank Johannes Henn for many useful hints in connection to
the construction of the canonical basis.  The work of WBK is supported by the
U.S. Department of Energy under Contract No. DE-AC02-98CH10886.  Parts of this
work were supported by the European Commission through contract
PITN-GA-2012-316704 (HIGGSTOOLS), by BMBF through Grant No. 05H12VKE, 
and by the ERC Advanced Grant no. 320651 ``HEPGAME''.

%- }}}

%- {{{ bibliography

%- }}}

\end{document}